\documentclass{basi}
\usepackage{txfonts}
\usepackage{epsfig}
\def\bi{\bibitem{}}

\def\beb{\begin{thebibliography}{10}}
\def\eeb{\end{thebibliography}}
\def\bei{\begin{itemize}}
\def\eei{\end{itemize}}
\def\bef{\begin{figure}}
\def\eef{\end{figure}}
\def\ben{\begin{enumerate}}
\def\een{\end{enumerate}}
\def\beq{\begin{equation}}
\def\eeq{\end{equation}}
\def\ber{\begin{eqnarray}}
\def\eer{\end{eqnarray}}
\def\edo{\end{document}}      

\def\Bb{{\bf B}}
\def\pa{\partial}
\def\vb{{\bf v}}

\def\half{\frac{1}{2}}
\def\third{\frac{1}{3}}

\newcommand{\mdot}{\mbox{$\dot{M}$}}

\begin{document}
\title[Magnetic Field of Neutron Stars]{The Binary History and the
Magnetic Field of Neutron Stars}
\author[Sushan Konar]%
       {Sushan Konar\thanks{e-mail:sushan@cts.iitkgp.ernet.in} \\ 
        i. Department of Physics \& Meteorology, 
        ii. Centre for Theoretical Studies, \\
        Indian Institute of Technology, Kharagpur 721302}
\maketitle
\label{firstpage}
\begin{abstract}
There has been strong observational evidence suggesting a causal 
connection between the binary history of neutron stars and the 
evolution of their magnetic field. In this article we discuss one
of the plausible mechanisms proposed for the evolution of the 
surface magnetic field, that of the diamagnetic screening of the 
field by accreted material. 
\end{abstract}


\section{Introduction}
\label{sintro}

Radio  pulsar   observations  have  established   that  neutron  stars
associated  with binaries have  magnetic fields  significantly smaller
than in  isolated neutron stars  which have field  strengths clustered
around  $10^{12}$~G  (excluding  magnetars).  It  is  understood  that
processing in  binaries alter the  magnetic field strength  of neutron
stars   giving  rise   to   the  low-field   binary  and   millisecond
pulsars.  Unfortunately,  till  date   there  is  no  clear  consensus
regarding either the nature of  the internal configuration or the time
evolution of  the magnetic  field in neutron  stars. Depending  on the
generation  mechanism  the field  could  either  be  supported by  the
crustal  currents   or  by  the  Abrikosov  fluxoids   of  the  proton
superconductor in the stellar core.

Accordingly,  two  classes  of  models  have  been  proposed  for  the
evolution  of the  magnetic field  in  accreting neutron  stars -  one
relating the  magnetic field  evolution to the  spin evolution  of the
star  assuming  the  field  to  be contained  in  the  superconducting
fluxoids  and the  other  attributing the  field  evolution to  direct
effects of  mass accretion  on the crustal  currents. In  an accretion
heated crust, the  decay takes place principally as  a result of rapid
dissipation  of  currents  due  to  the  decrease  in  the  electrical
conductivity  and hence  a  reduction in  the  ohmic dissipation  time
scale.  Interestingly, the  mechanism of  ohmic decay,  unique  to the
crustal currents,  is also used  in models where spin-down  is invoked
for flux expulsion,  for a subsequent dissipation of  such flux in the
crust. Therefore,  the theoretical  efforts in modeling  the plausible
mechanisms have so  far been concentrated on the  ohmic dissipation of
currents in the outer  crust (Konar \& Bhattacharya 2001; Bhattacharya
2002).

Another  possible mechanism  is to  screen  the surface  field by  the
accreting material.  As the  highly conducting accreting  plasma flows
horizontally  from the  polar caps  to lower  latitudes,  the magnetic
field  lines  are  dragged along  with  it,  by  virtue of  {\it  flux
freezing}.  This  dragging may  lead  to  the  creation of  additional
horizontal  components at the  expense of  vertical ones  producing an
effective {\it screening} of the dipolar surface field. Interestingly,
this mechanism  would not depend on  the location of the  field in the
stellar interior and would be  effective irrespective of the nature of
the interior currents.  Even though this was suggested  quite early on
(Bisnovatyi-Kogan \&  Komberg 1974;  Blandford, De Campli  \& K\"onigl
1979; Taam  \& van  den Heuvel  1986; Romani 1990,  1995), it  is only
recently  that  the  problem  is being  investigated  in  quantitative
detail. One-dimensional plane-parallel modeling by Cumming, Zweibel \&
Bildsten   (2001)  indicates   that  the   diamagnetic   screening  is
ineffective  for  field  strengths   above  $\sim  10^{10}$G  and  for
accretion   rates  below   $\sim$  1   \%  of   the   local  Eddington
rate.  Recently,  Melatos  \&   Phinney  (2001)  have  calculated  the
hydromagnetic structure  of a neutron star  accreting symmetrically at
both magnetic poles  as a function of the  accreted mass. According to
their  calculation the  magnetic  dipole moment  scales as  $B_0^{1.3}
\mdot^{0.18}  M_a^{\-1.3}$  where $B_0$,  $\mdot$  and  $M_a$ are  the
initial field strength,  the rate of accretion and  the total accreted
mass.

\section{Material Flow : 2-Dimensional Model}

\bef
\begin{center}{
\epsfig{file=fig01_a.ps,width=185pt,angle=-90}
\hspace{6.85cm}
\epsfig{file=fig01_b.ps,width=185pt,angle=-90}}
\end{center}
\caption[] {Flow velocity (marked $a$), $\rho \vb$, and its divergence
(marked $b$)  in a  right-angular slice ($0  \leq \theta  \leq \pi/2$,
$0.25 \leq  r \leq 1.0$)  with $r_m =  0.75$, $r_b = 0.5$.  The panels
marked $1, 2$ correspond to $c = 0.0, 1.0$.}
\label{fig01}
\eef          

In a recent work we have  proposed a 2-dimensional flow pattern of the
accreted  material   to  demonstrate  the   mechanism  of  diamagnetic
screening (Choudhuri  \& Konar 2002). The  accreted material, confined
to the poles by strong  magnetic stresses, accumulates in a column and
sinks  below the  surface when  the pressure  of the  accretion column
exceeds  the  magnetic pressure.  From  the  bottom  of the  accretion
column, the material  from both the poles move  to the lower latitudes
in  an {\em  equator-ward} flow,  meet  at the  equator and  submerge,
pushing against the solid interior  and displacing it very slowly in a
{\em counter-flow} down-wards as well as to higher latitudes. Finally,
in the very deep layers the material moves radially in-wards due to an
overall compression  of the  star. In the  top layer  $r_m < r  < r_s$
($r_s$ - stellar radius) of equator-ward flow, we have :
\ber
\rho v_r^{\bf 0} 
&=& K_1 \left[\frac{1}{3} r - \frac{1}{2} r_m 
  + \frac{\left( \frac{1}{2} r_m 
  - \frac{1}{3} r_s \right) r_s^2}{r^2} \right] e^{- \beta \cos^2 \theta} \,,  \\
\rho v_{\theta}^{\bf 0} 
&=& \half \sqrt{\frac{\pi}{\beta}} K_1 \frac{(r - r_m)}{\sin \theta} 
  \mbox{erf} (\sqrt{\beta} \cos \theta) \left(1 - e^{-\gamma \theta^2} \right) \,. 
\eer
In the layer beneath, $r_b < r <r_m$, where the flow turns around, we have :
\ber
\rho v_r^{\bf 0} 
&=& K_2 \, e^{-\beta \cos^2 \theta} 
   \left[\third r - \half (r_m + r_b) + \frac{r_m r_b}{r} 
   + \frac{\left(\frac{1}{6} r_b - \half r_m \right) r_b^2}{r^2} \right] \nonumber \\
&-& \half \sqrt{\frac{\pi}{\beta}} K_2 \,
    \left[\third r - \half (r_m + r_b) + \frac{r_m r_b}{r} 
    + \frac{\left( \frac{1}{6} r_m - \half r_b \right) r_m^2}{r^2} \right]  \,,\\
\rho v_{\theta}^{\bf 0} 
&=& \frac{K_2}{2 \sin \theta} \sqrt{\frac{\pi}{\beta}} 
  \left( r + \frac{r_m r_b}{r} - r_b - r_m \right) 
  \left[\mbox{erf}(\sqrt{\beta} \cos \theta) - \cos \theta \right] \,.
\eer
Finally, when $r<r_b$, the velocity is radially inward (characteristic of 
the radial compression in the deeper layers): 
\ber
\rho v_r^{\bf 0} = - \frac{K_3}{r^2}\,, \; \; \; \; \mbox{and} \; \; \; \;
\rho v_{\theta}^{\bf 0} = 0 \,.
\eer
It is evident  from the context that the rate  of accretion is related
to  these coefficients  by the  relation  $K_3 =  \mdot/{4 \pi}$.  The
coefficients $K_1$, $K_2$  and $K_3$ are related due  to the fact that
$\rho v_r$ has to be continuous across $r = r_m$ and $r = r_b$. Notice
that whereas $\gamma$  defines the size of the  polar cap (the angular
extent being  given by  $\sim \gamma^{-1/2}$), $\beta$  determines the
magnitude  of   the  material  flow   that  sinks  inward   below  the
equator. The  new material enters our  region of interest  only at the
polar cap and the divergence in the top layer ($r \ge r_m$) is :
\ber
\nabla. (\rho \vb^{\bf 0}) 
= K_1 \, e^{-\gamma \theta^2 - \beta \cos^2\theta} \, \frac{r - r_m}{r} 
  \times \left(1 + \sqrt{\frac{\pi}{\beta}} \frac{\gamma\theta}{\sin \theta}
   e^{\beta \cos^2\theta} \mbox{erf}(\sqrt{\beta} \cos\theta) \right)
\eer
providing  for a  source of  material  only around  the polar  region.
$\nabla.(\rho \vb) = 0$ everywhere else.

However, with a  decrease in the magnetic field  the magnetic pressure
in the polar  region reduces, allowing the inflow  of material through
an  increasingly  larger  region  around  the  pole.  Eventually,  the
magnetic field becomes too small to  be able to channelize the flow of
material  and   the  accretion  becomes   spherically  symmetric.   We
represent this  effect through  a parameter $0  \leq K_0 \leq  1$. The
effect  of the widening  of the  polar cap  is introduced  through the
relation $\gamma = (\theta_{\rm  min} + K_0 \Delta \theta)^{-2}$ where
$\theta_{\rm  min}$ is  the angular  extent  of the  polar cap  before
accretion begins.  In order to  make the flow velocity  more isotropic
with the widening of the polar cap, an isotropic part is also added to
the velocity field. Therefore,  the expression for the velocity field,
for   a   time-dependent  magnetic   field   strength   and  hence   a
time-dependent polar cap area, is given by :
\beq
\vb = (1-K_0) \, \vb^{\bf 0}\left(\gamma(c)\right) + K_0 \vb^{\bf 1} 
\label{eq_vmod} \,,
\eeq
where $\vb^{\bf 1}$ is the purely isotropic part : 
$ \rho \vb^{\bf 1} 
= -\frac{K_3}{r^2} 
  \left( 1 - \exp \left(5 \frac{r_s - r}{r_s - r_m} \right) \right)\,$.
Relating the extent of the polar cap with the surface field strength
at a given instant of time (Konar 1997) we find that
\beq
\frac{\sin (\theta_{\rm min} + K_0(t) \, \Delta \theta) }{\sin \theta_{\rm min}} 
= \left[ \frac{B_s(t)}{B_s(t=0)} \right]^{-2/7}\,. 
\eeq
It  should be  noted that  this relation  is valid  till  accretion is
purely spherical and $K_0$ equals  unity. Fig.[1] shows the profile of
the velocity field and its divergence for different values of $K_0$.

\section[]{Evolution of the Magnetic Field}

The  magnetic field  of  the  neutron star  evolves  according to  the
induction equation:
\beq
\frac{\pa \Bb}{\pa t} 
= \nabla \times (\vb \times \Bb) 
  - \frac{c^2}{4\pi} \nabla \times (\frac{1}{\sigma} \nabla \times \Bb) \,,
\eeq
where $\sigma$ is the  electrical conductivity of the medium. Assuming
an axisymmetric poloidal field,  allowing us to represent the magnetic
field of the form, $\Bb  = \nabla \times \left( A(r, \theta) \hat{{\bf
e_\phi}}\right)$, we find that $A$ evolves according to the equation:
\beq
\frac{\pa A}{\pa t} + \frac{1}{s} (\vb. \nabla)(s A) 
= \eta \left( \nabla^2 - \frac{1}{s^2} \right)A \,, 
\label{eq_dadt}
\eeq
where $\eta =  c^2/4\pi\sigma$ and $s = r  \sin \theta$. Evidently, it
is  the poloidal  component of  $\vb$  that affects  the evolution  of
$A$.  The evolution of  the magnetic  field with  time is  obtained by
integrating   equation  (\ref{eq_dadt})   subject   to  the   boundary
conditions that the field lines  from the two hemispheres should match
smoothly at the equator, requiring $\partial A/\partial \theta = 0$ at
$\theta =  \pi/2$ and $A =  0$ at $\theta =  0$ such that  there is no
singularity at the pole.

\bef
\begin{center}
\epsfig{file=fig02_a.ps,width=180pt}
\end{center}
\begin{center}
\epsfig{file=fig02_b.ps,width=140pt,angle=-90}
\end{center}
\caption[]{Evolution  of  the mid-latitude  surface  field with  time.
$Top$ - For a steady flow pattern with $\eta=0.01$. The curves 1 and 2
corresponds to cases without and with magnetic buoyancy (characterized
by a  radially outward velocity in  the top layer taken  to be $v_{\rm
mb} =  50$ here).  $Bottom$ -  For a time-dependent  flow pattern with
$\eta=0.05$ and $v_{\rm mb} = 50.0$.}
\label{fig02}
\eef

Our calculation  indicates that the  surface field is screened  in the
short time scale of the flow of accreting material in the top layer if
magnetic  buoyancy  is neglected,  whereas  on  inclusion of  magnetic
buoyancy the screening  takes place in the somewhat  longer time scale
of  the slow  interior flows,  as  seen if  Fig.[2a]. Since,  magnetic
buoyancy is expected  to be important in the  liquid surface layers of
accreting neutron stars, it is  the second time-scale, estimated to be
of the order of $10^5$ years, which is of real importance. Remarkably,
this is  comparable to  the time-scale of  accretion in  massive X-ray
binaries indicating that the diamagnetic screening could indeed be one
of the viable mechanisms of field reduction.

\bef
\begin{center}{
\epsfig{file=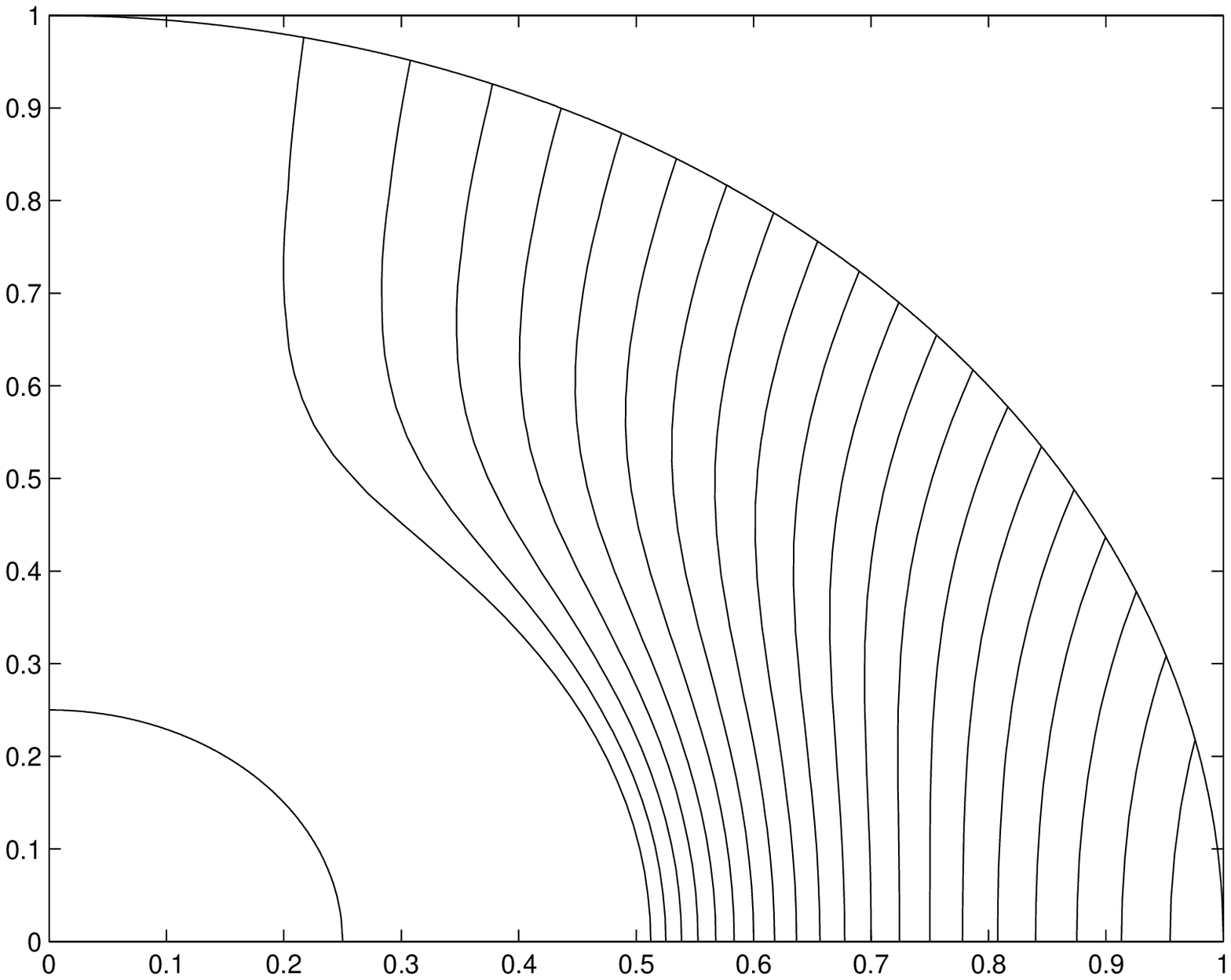,width=140pt}
\epsfig{file=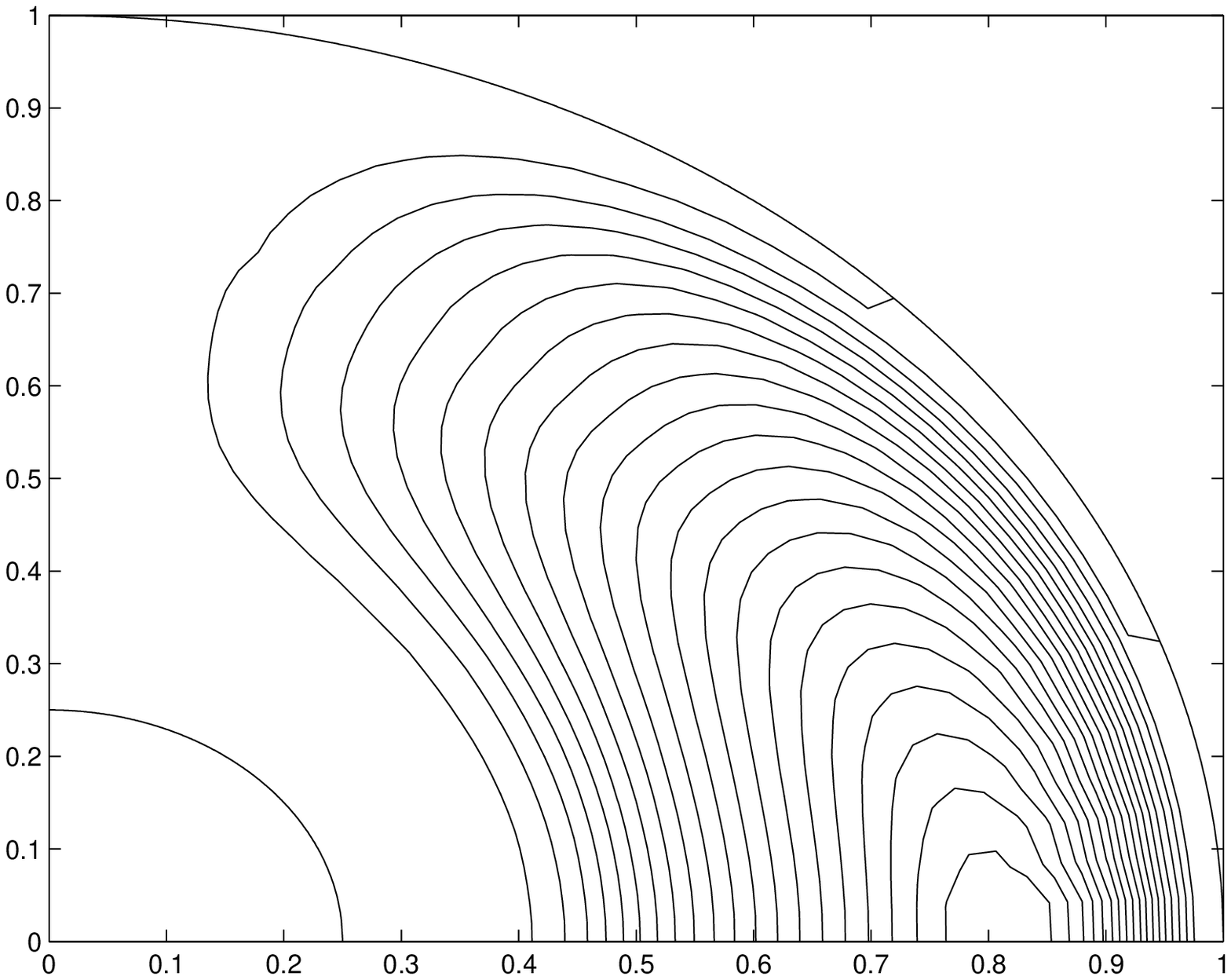,width=140pt}
}
\end{center}
\begin{center}{
\epsfig{file=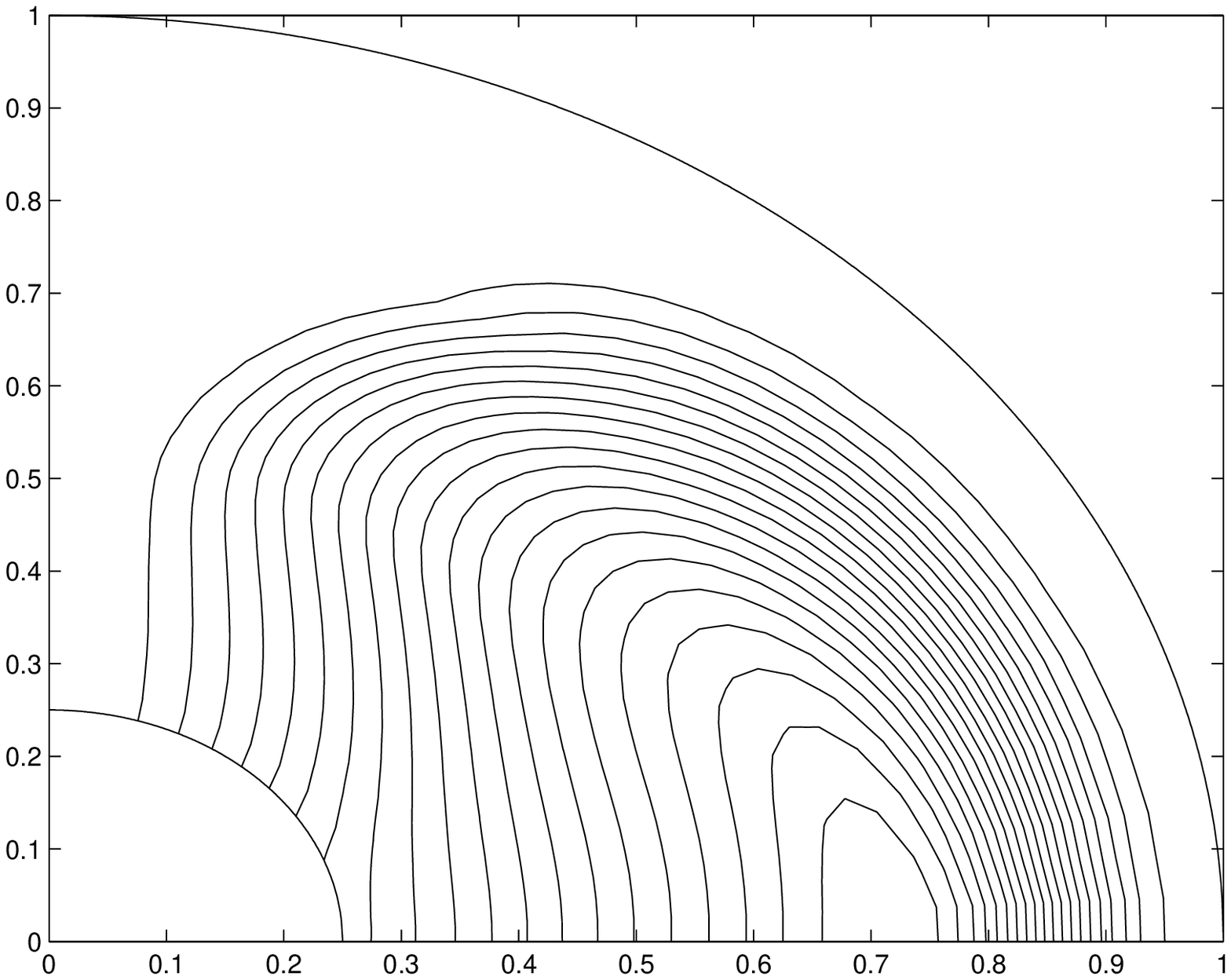,width=140pt}
\epsfig{file=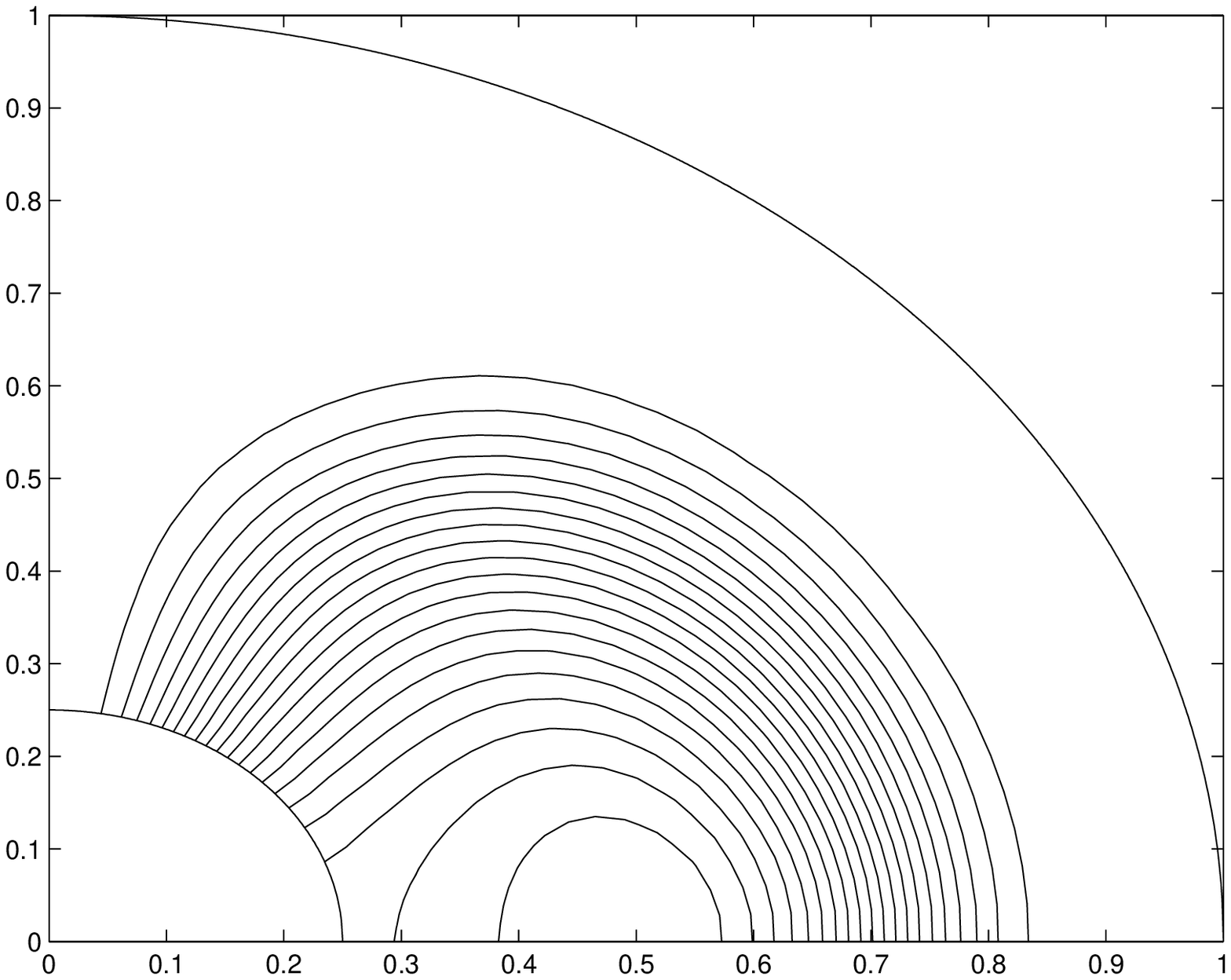,width=140pt}
}
\end{center}
\caption[]{Field  configurations  corresponding  to  bottom  panel  of
figure  {\bf  2}, at  intermediate  times  starting  from the  initial
configuration of  top left-hand panel.   The panels top  right, bottom
left  and bottom right  panels correspond  to $t  = 0.015,  0.05, 0.1$
respectively.}
\label{fig03}
\eef

The  most interesting  conclusion of  our work  is the  fact  that the
screening  becomes progressively  less effective  with  the decreasing
strength of the surface field as  seen if Fig.[2b] and Fig.[3]. As the
magnetic field  at the surface decreases  due to screening,  it can no
longer channelize the material flow and the accretion becomes more and
more  spherical.  Magnetic  field  on  the surface  of  a  spherically
accreting   star   is  not   subject   to   dragging  and   subsequent
burial. Therefore,  after an initial  rapid decrease, the  decay slows
down  and the  field  reaches  an asymptotic  value  as the  accretion
becomes completely spherical. It should be mentioned here that similar
conclusions were drawn by Konar \& Bhattacharya (1997) for the case of
crustal currents undergoing  accretion-induced ohmic dissipation where
a purely spherical accretion was assumed to be operative at all times,
even though  the detailed crustal micro-physics,  responsible for such
behaviour,    has   not   been    incorporated   in    the   screening
model. Therefore, further investigation  is needed to be undertaken by
combining the  effects of screening and ohmic  dissipation taking into
account the  detailed micro-physics of  the neutron star  crust (Konar
2003).

Another important  aspect that  requires further investigation  is the
question of  the evolution  of the screened  magnetic field  after the
cessation of  accretion. The effect of magnetic  buoyancy, causing the
field to  re-emerge to  the surface would  be significant only  if the
flux  resides   in  the   topmost  liquid  layers.   Similarly,  ohmic
dissipation  of the  field would  be most  effective in  the outermost
layers by virtue  of smaller conductivity. It has  been seen (Konar \&
Bhattacharya 1999a,  199b) that for purely spherical  accretion by the
time accretion ceases, for almost all realistic binary parameters, any
initial current  configuration (crustal or core) would  be buried deep
inside    the   core    of   the    star   preventing    any   further
evolution. Therefore,  we do not expect any  effective re-emergence of
the field. However detailed quantitative estimates are still pending.

The  details of the  screening model  discussed here  can be  found in
Choudhuri \& Konar (2002) and Konar \& Choudhuri (2002, 2003).

\section*{Acknowledgements}

The  work presented  here has  been done  in collaboration  with Arnab
R. Choudhuri.  Discussions with Dipankar  Bhattacharya, Denis Konenkov
and U.~R.~M.~E.  Geppert have  been extremely useful.  A post-doctoral
fellowship at  IUCAA, Pune and hospitality provided  by the department
of  Physics,  IISc,  Bangalore   for  my  periodic  visits  have  been
instrumental for the completion of this work.

\beb
\bi Bhattacharya D., 2002, \newblock {\it JAA}, {\bf 22}, 67
\bi Bisnovatyi-Kogan G.~S., Komberg B.~V., 1974, \newblock {\it SvA}, {\bf 18}, 217
\bi Blandford R.~D., DeCampli W.~M., K\"onigl A., 1979, \newblock {\it BAAS}, {\bf 11}, 703
\bi Choudhuri, A.~R. and Konar, S., 2002, \newblock {\it MNRAS}, {\bf 332}, 933
\bi Cumming A., Zweibel E., Bildsten L., 2001, \newblock {\it ApJ}, {\bf 557}, 958
\bi Konar S., 1997, \newblock {\it PhD Thesis}, IISc, Bangalore
\bi Konar S., Bhattacharya D., 1997, \newblock {\it MNRAS}, {\bf 284}, 311
\bi Konar S., Bhattacharya D., 1999a, \newblock {\it MNRAS}, {\bf 303}, 588
\bi Konar S., Bhattacharya D., 1999b, \newblock {\it MNRAS}, {\bf 308}, 795
\bi Konar S., Bhattacharya D., 2001, \newblock In Kouveliotou C., van Paradijs J., 
    Ventura J., editors {\it The Neutron Star - Black Hole Connection, NATO Science 
    Series C, Vol. 267}, page~71, Kluwer Academic Publishers 
\bi Konar, S. and Choudhuri, A.~R., 2002, \newblock {\it BASI}, {\bf 30}, 697
\bi Konar, S. and Choudhuri, A.~R., 2003, \newblock {\tt astro-ph/0304490}
\bi Konar, S., 2003, \newblock {\it in preparation}
\bi Melatos A., Phinney E.~S., 2001, \newblock {\it PASP}, {\bf 18}, 421
\bi Romani R. ~W., 1990, \newblock {\it Nat}, {\bf 347}, 741
\bi Romani R. ~W., 1995, \newblock In van Riper K., Epstein R., Ho C., editors 
    {\it Isolated Pulsars}, page~75, Cambridge University Press
\bi Taam R.~E., van den Heuvel E.~P.~J., 1986, \newblock {\it ApJ}, {\bf 305}, 235
\eeb

\label{lastpage}
\end{document}